\documentstyle{article}
\begin{document}

\hoffset -1.7cm

\setlength{\hsize}{14.8cm}
\setlength{\vsize}{20cm}

\title{\bf Systematic Study of 3-3-1 Models}
\author{{\bf William A. Ponce and Yithsbey Giraldo}\\
{\it Instituto de F\'\i sica, Universidad de Antioquia,}\\
{\it A.A. 1226, Medell\'\i n, Colombia.}
\and
{\bf Luis A. S\'anchez}\\
{\it Escuela de F\'\i sica, Universidad Nacional de Colombia}\\
{\it A.A. 3840, Medell\'\i n, Colombia.}}

\date{}
\maketitle

\begin{abstract}

We carry a systematic study of possible models based on the local gauge
group $SU(3)_c\otimes SU(3)_L\otimes U(1)_X$. Old and new models emerge
from the analysis.

\end{abstract}

\vspace{0.5cm}
\large
The Standard Model (SM) based on the local gauge group $SU(3)_c\otimes
SU(2)_L\otimes U(1)_Y$ \cite{sm} can be extended in different ways:
first, by adding new fermion fields (adding a right-handed
neutrino field constitute its simplest extension, and has deep
consequences as for example, the implementation of the see-saw mechanism,
and the enlarging of the possible number of local
abelian symmetries that can be gauged simultaneously); second, by
augmenting the scalar sector to more than one higgs representation, and
third by enlarging the local gauge group. In this last direction,
$SU(3)_L\otimes U(1)_X$ as a flavor group has been studied previously by
several authors in the literature \cite{fp}-\cite{mps} who have explored
possible fermion and higgs-boson representation assignments.

In what follows we are going to present a systematic analysis of the local
gauge model based on the gauge structure $SU(3)_c\otimes
SU(3)_L\otimes U(1)_X$, which we call the 331 theory.

We assume that the electroweak group is $SU(3)_L\otimes
U(1)_X\supset SU(2)_L \otimes U(1)_Y$. We also assume that the left handed
quarks (color triplets), left-handed leptons (color singlets) and scalars, transform
under the two fundamental representations of $SU(3)_L$ (the 3 and $3^*$).
Two classes of models will be discussed: one family models where the
anomalies cancel in each family as in the SM, and family models where the
anomalies cancel by an interplay between the families. $SU(3)_c$ is vectorlike as in the SM.

The most general expression for the electric charge generator in
$SU(3)_L\otimes U(1)_X$ is a linear combination of the three diagonal
generators of the gauge group
\begin{equation}\label{ch}
Q=aT_{3L}+\frac{2}{\sqrt{3}}bT_{8L}+xI_3,
\end{equation}

\noindent
where $T_{iL}=\lambda_{iL}/2$, being $\lambda_{iL}$ the Gell-Mann matrices
for $SU(3)_L$ normalized as {\bf Tr}$(\lambda_i\lambda_j)=2\delta_{ij}$,
$I_3=Dg.(1,1,1)$ is the diagonal $3\times 3$ unit matrix,
and $a$ and $b$ are arbitrary parameters to be determined anon. Notice
that we can absorb an eventual coefficient for $x$ in its definition.

If we assume that the usual isospin $SU(2)_L$ of the SM is such that
$SU(2)_L\subset SU(3)_L$, then $a=1$ and we have just one parameter set
of models, all of them characterized by the value of $b$. So, Eq.
(\ref{ch}) allows for an infinite number of models in the context of the
331 theory, each one associated to a particular value of the
parameter $b$, with characteristic signatures that
make one different from other, as we will see.

There are a total of 17 gauge bosons in the gauge group under
consideration, they are: one gauge
field $B^\mu$ associated with $U(1)_X$, the 8 gluon fields
associated with $SU(3)_c$ which remain massless after breaking the
symmetry, and another 8 associated
with $SU(3)_L$ that we may write in the following way:

\[{1\over 2}\lambda_{\alpha L} A_\mu^\alpha={1\over \sqrt{2}}\left(
\begin{array}{ccc}D^0_{1\mu} & W^+_\mu & K^{(1/2+b)}_\mu \\ W^-_\mu &
D^0_{2\mu} & K^{-(1/2-b)}_\mu \\ K^{-(1/2+b)}_\mu & K^{(1/2-b)}_\mu &
D^0_{3\mu} \end{array}\right), \]
where $D^\mu_1=A_3^\mu/\sqrt{2}+A_8^\mu/\sqrt{6},\;
D^\mu_2=-A_3^\mu/\sqrt{2}+A_8^\mu/\sqrt{6}$, and
$D^\mu_3=-2A_8^\mu/\sqrt{6}$. The upper indices in the gauge bosons in
the former expression
stand for the electric charge of the corresponding particle, some of them
functions of the $b$ parameter as they should be.  Notice that the gauge
bosons have integer electric charges only for $b=\pm 1/2,\; \pm 3/2, ; \pm
5/2,...,\pm (2n+1)/2, \; n=1,2,3...$. A deeper analysis shows that the
negative values for $b$ can be related to the positive one just by taking
the complex conjugate in the covariant derivative of each model, which in
turn is equivalent to replace $3\leftrightarrow 3^*$ in the fermion
content of each particular model. So, our first conclusion is that, if we
do not want exotic electric charges in the gauge sector of our theory,
then $b$ must be equal to 1/2.  We will see next that this is also
the condition for excluding exotic electric charges in the fermion sector.

Now, contrary to the SM where only the abelian $U(1)_Y$ factor is
anomalous, in the 331 theory both, $SU(3)_L$ and $U(1)_X$ are anomalous
($SU(3)_c$ is vectorlike). So, special combination of multiplets must be
used in each particular model in order to cancel the possible
anomalies, and end with renormalizable models. The triangle anomalies
we must take care of are: $[SU(3)_L]^3$,
$[SU(3)_c]^2U(1)_X$, $[SU(3)_L]^2U(1)_X$, $[grav]^2U(1)_X$ and
$[U(1)_X]^3$.

In order to present specific examples, let us see how the charge
operator in Eq.(\ref{ch}) acts on the representations 3 and $3^*$ of
$SU(3)_L$:

\begin{eqnarray*}
Q[3]
&=&Dg.({1\over 2}+{b\over 3}+x, -{1\over 2}+{b\over 3}+x, -{2b\over
3}+x)\\
Q[3^*]
&=&Dg.(-{1\over 2}-{b\over 3}+x, {1\over 2}-{b\over 3}+x, {2b\over 3}+x).
\end{eqnarray*}

\noindent
Notice from the former expressions that, if we accommodate the known
left-handed quark and lepton isodoublets in the two upper components of 3
and $3^*$ (or $3^*$ and 3), and forbid the presence of exotic electric
charges in the possible models, then the electric charge of the third
component in those representations must be equal either to the charge of
the first or second component, which in turn implies $b=\pm 1/2$. Since the
negative value is equivalent to the positive one, $b=1/2$ is a necessary
and sufficient condition in order to exclude exotic electric charges in
the fermion sector.

\section{Some examples}
\subsection{The Pleitez-Frampton model}
As a first example let us take $b=3/2$, consecuently
$Q[3]=Dg.(1+x,x,-1+x)$ and
$Q[3^*]=Dg.(-1+x,x,1+x)$. Then the following multiplets are associated
with the respective $(SU(3)_c,SU(3)_L,U(1)_x)$ quantun numbers:
$(e^-,\nu_e,e^+)^T_L\sim (1,3^*,0); \;\; (u,d,j)^T_L\sim (3,3,-1/3)$
and $(d,u,s)^T_L\sim (3,3^*,2/3)$, where $j$ and $s$ are isosinglets
exotic quarks of electric charges $-4/3$ and 5/3 respectively. This
multiplet structure is the basis of the Pleitez-Frampton
model \cite{fp} for which the anomaly-free arrangement for three
families is given by:

\begin{eqnarray*}
\psi_L^a&=& (e^a,\nu^a,e^{ca})^T_L\sim (1,3^*,0)\\
q_L^i&=&(u^i,d^i,j^i)_L^T\sim (3,3,-1/3)\\
q_L^1&=&(d^1,u^1,s)_L^T\sim (3,3^*,2/3)\\
u_L^{ca}&\sim & (3,1,-2/3),\;\; d_L^{ca}\sim (3,1,1/3)\\
s_L^c&\sim & (3,1,-5/3),\;\; j_L^{ci}\sim (3,1,-4/3),
\end{eqnarray*}

\noindent
where the upper $c$ symbol stands for charge conjugation,
$a=1,2,3$ is a family index and $i=2,3$ is related to two of the 3
families (in the 331 basis). As can be seen, there are six triplets of
$SU(3)_L$ and six anti-triplets, which ensures cancellation of the
$[SU(3)_L]^3$ anomaly. A power counting shows that the other four
anomalies also vanish.

\subsection{Other 331 family models in the literature}
Let us analyze other two 331 three family models already present in the
literature, for which $b=1/2$ (they do not contain
exotic electric charges). For that
particular value of $b$ we have: $Q[3]=Dg. (2/3+x,-1/3+x,-1/3+x)$ and
$Q[3^*]=Dg.(-2/3+x,1/3+x,1/3+x)$. Then we get the following multiplets
associated with the given quantun numbers: $(u,d,D)^T_L\sim (3,3,0),\;\;
(e^-,\nu_e,N^0)^T_L\sim (1,3^*,-1/3)$ and $(d,u,U)^T_L\sim
(3,3^*,1/3)$, where $D$ and $U$ are exotic quarks with electric charges
$-1/3$ and 2/3 respectively. With this gauge structure we may construct
the following anomaly free model for three families:

\begin{eqnarray*}
\psi _L^{'a}&=& (e^a,\nu^a,N^{0a})^T_L\sim (1,3^*,-1/3)\\
q_L^{'i}&=&(u^i,d^i,D^i)_L^T\sim (3,3,0)\\
q_L^{'1}&=&(d^1,u^1,U)_L^T\sim (3,3^*,1/3)\\
u_L^{ca}&\sim & (3,1,-2/3),\;\; d_L^{ca}\sim (3,1,1/3)\\
U_L^c&\sim & (3,1,-2/3),\;\; D_L^{ci}\sim (3,1,1/3),
\end{eqnarray*}

\noindent
where $a=1,2,3$ is a family index and  $i=2,3$.
This model has been analyzed in the literature in Ref. \cite{long}. If
needed, this model can be augmented with an undetermined number of
neutral Weyl states $N^{0b}_L\sim (1,1,0),\; b=1,2,...$, without
violating the anomaly constraint relations. We call this {\bf Model A}.

The other model has the same quark multiplets used in the previous model
arranged in a different way, and it makes use of a new lepton
multiplet $\psi"_L=(\nu_e,e^-,E^-)^T_L\sim (1,3,-2/3)$. The
multiplet structure of this new anomaly-free three family model is given
by:

\begin{eqnarray*}
\psi "_L^a&=& (\nu^a,e^a,E^a)^T_L\sim (1,3,-2/3)\\
e^{ca}&\sim & (1,1,1),\;\; E^{ca}\sim (1,1,1)\\
q"_L^1&=&(u^1,d^1,D)_L^T\sim (3,3,0)\\
q"_L^i&=&(d^i,u^i,U^i)_L^T\sim (3,3^*,1/3)\\
u_L^{ca}&\sim & (3,1,-2/3),\;\; d_L^{ca}\sim (3,1,1/3)\\
D_L^c&\sim & (3,1,1/3),\;\; U_L^{ci}\sim (3,1,2/3).
\end{eqnarray*}

\noindent
This model has been analyzed in the literature in Ref. \cite{ozer}. We
call this {\bf Model B}.

\subsection{Other models}
Now we want to consider other possible 331 models without exotic electric
charges ($b=1/2$). Let us start first
defining the following closed set of fermions (closed in the
sense that they include the antiparticles of the charged particles):\\
$S_1=[(\nu_\alpha, \alpha^- ,E_\alpha^-); \alpha^+;
E_\alpha^+]$ with
quantum numbers $[(1,3,-2/3); (1,1,1); (1,1,1)]$.\\
$S_2=[(\alpha^- , \nu_\alpha ,N_\alpha^0); \alpha^+]$ with
quantum numbers $[(1,3^*,-1/3); (1,1,1)]$.\\
$S_3=[(d,u,U);u^c;d^c;U^c]$ with
quantum numbers $(3,3^*,1/3); (3^*,1,-2/3);(3^*,1,1/3)$ and
$(3^*,1,-2/3)$ respectively.\\
$S_4=[(u,d,D); d^c;u^c;D^c]$ with
quantum numbers $(3,3,0); (3^*,1,1/3); (3^*,1,-2/3)$ and
$(3^*,1,1/3)$ respectively.\\
$S_5=[(e^-,\nu_e ,N_1^0); (E^-,N_2^0,N_3^0)$; $(N_4^0,E^+,e^+)]$ with
quantum numbers $(1,3^*,-1/3)$; $(1,3^*,-1/3)$ and
$(1,3^*,2/3)$ respectively.\\
$S_6=[(\nu_e, e^-, E^-); (E_2^+, N_1^0,N_2^0);(N_3^0, E_2^- ,E_3^-);
e^+;E_1^+;E_3^+]$ with quantum numbers
[$(1,3,-2/3)$; $(1,3,1/3)$; $(1,3,-2/3)$; $(1,1,1)$; $(1,1,1)$; $(1,1,1)$]

The anomalies for the former sets are presented in the following Table.

\begin{center}

TABLE I. Anomalies for $S_i$.

\begin{tabular}{||l||c|c|c|c|c|c||}\hline\hline
Anomalies           & $S_1$& $S_2$& $S_3$& $S_4$& $S_5$ & $S_6$ \\
\hline\hline
$[SU(3)_c]^2U(1)_X$ & 0    &  0    &  0    &  0   &  0  & 0 \\
$[SU(3)_L]^2U(1)_X$ & $-2/3$ & $ -1/3$ &  1    &  0   &  0  & $-1$ \\
$[grav]^2U(1)_X$     & 0    &  0    &  0    &  0   &  0  & 0 \\
$[U(1)_X]^3$        & 10/9 & 8/9   & $-12/9$ & $-6/9$ & 6/9 & 12/9 \\
$[SU(3)_L]^3$       & 1    & -1 & -3 & 3 & -3 & 3 \\
\hline\hline
\end{tabular}
\end{center}

\vspace{.5cm}

\noindent
Notice from the Table that {\bf Model A} is given by
$(3S_2+S_3+ 2S_4)$ and {\bf Model B} by $(3S_1+2S_3+S_4)$. But they are
not the only anomaly-free structures we may build. Let us see:

\subsubsection{One family models}
There are two anomaly-free one family structures that can be extracted
from the Table. They are:\\
{\bf Model C}: $(S_4+S_5)$. This model is associated with an $E_6$
subgroup and has been analyzed in Ref. \cite{spm}.\\
{\bf Model D}: $(S_3+S_6)$. This model is associated with an
$SU(6)_L\otimes U(1)_X$ subgroup and has been analyzed in Ref. \cite{mps}.

The former two models can become realistic models (for 3 families) just
by carbon copy each family as in the SM, that is, taking $3(S_4+S_5)$ and
$3(S_3+S_6)$.

\subsubsection{Two family models}
There are three two family models. They are given by: $(S_1+S_2+S_3+S_4)$,
$2(S_4+S_5)$ and $2(S_3+S_6)$. These three models are not
realistic.

\subsubsection{Three family models}
Besides models {\bf A} and {\bf B} we have two more. They are:
{\bf Model E}: $(S_1+S_2+S_3+2S_4+S_5)$ and  {\bf Model F}:
$(S_1+S_2+2S_3+S_4+S_6)$.\\

The main feature of these last two models is that, contrary to
all the other models, each one of the three families is treated in a
different way. As far as we know, these two models have not been studied
in the literature so far.

We may construct now four, five, etc. family models (a four family
model is given for example by: $2(S_1+S_2+S_3+S_4)$), but as for the
two family case, they are not realistic.

\section{The scalar sector}
Even though the representation content for the fermion fields may vary
significantly from model to model, all $SU(3)_L\otimes U(1)_X$ models
presented have a gauge boson sector which depends only on the $b$
parameter. In what follows we are going to reffer only to models for
which $b=1/2$ (models {\bf A-F}). For that particular value of $b$ there
are only two Higgs scalars which may develop a nonzero Vacuum Expectation
Value (VEV), they are
$\phi_1(1,3^*,-1/3)=(\phi_1^-,\phi_1^0,\phi_1^{'0})$, with (VEV)
$\langle\phi_1\rangle=(0,v,V)^T$ and
$\phi_2(1,3^*,2/3)=(\phi_2^0,\phi_2^+\phi_2^{'+})$ with VEV
$\langle\phi_2\rangle=(v',0, 0)^T$, with the hierarchy $V>v\sim v'\sim
250$ GeV, the electroweak mass scale.

Our aim is to break the symmetry in the way

\[SU(3)_c\otimes SU(3)_L\otimes U(1)_X\longrightarrow
SU(3)_c\otimes SU(2)_L\otimes U(1)_Y\longrightarrow SU(3)_c\otimes
U(1)_Q,\]
and produce mass terms for the fermion fields at the same time.

In some models it is more convenient to work with a different set
of Higgs fields. For example, in the one family model in Ref. \cite{spm}
the following three scalars were used:
$\phi_1(1,3^*,-1/3)$ with $\langle\phi_1\rangle=(0,0,V)^T$,
$\phi_2(1,3^*,-1/3)$ with $\langle\phi_2\rangle=(0,v/\sqrt{2},0)^T$ and
$\phi_3(1,3^*,2/3)$ with $\langle\phi_3\rangle=(v'/\sqrt{2},0,0)^T$. For
that particular case we got the following mass terms for the charged gauge
bosons in the electroweak sector: $M^2_{W^\pm}=(g^2/4)(v^2+v^{'2}),\;
M^2_{K^\pm}=(g^2/4)(2V^2+v^{'2})$ and
$M^2_{K^0(\bar{K}^0)}=(g^2/4)(2V^2+v^2)$. For the neutral gauge
bosons we got a mass term of the form:
\[M=V^2(\frac{g'B^\mu}{3}-\frac{gA_8^\mu}{\sqrt{3}})^2
+ \frac{v^2}{8}(\frac{2g'B^\mu}{3}-gA^\mu_3 +\frac{gA_8^\mu}{\sqrt{3}})^2
+\frac{v'^2}{8}(gA_3^\mu-\frac{4g'B^\mu}{3}+\frac{gA^\mu_8}{\sqrt{3}})^2.\]

Diagonalizing $M$ and defining
\begin{eqnarray}\nonumber
Z_1^\mu&=&Z_\mu \cos\theta+Z'_\mu \sin\theta, \\ \nonumber
Z_2^\mu&=&-Z_\mu \sin\theta+Z'_\mu \cos\theta, \\ \nonumber
-\tan(2\theta)&=&\frac{\sqrt{12}C_W(1-T_W^2/3)^{1/2}[v'^2(1+T_W^2)-v^2(1-T_W^2)]}
{3(1-T_W^2/3)(v^2+v'^2)-C_W^2[8V^2+v^2(1-T_W^2)^2+v'^2(1+T_W^2)^2]},\\ \nonumber
\end{eqnarray}
we get that the photon field $A^\mu$ and the neutral fields $Z_\mu$ and
$Z'_\mu$ are given by
\begin{eqnarray} \nonumber
A^\mu&=&S_W A_3^\mu + C_W\left[\frac{T_W}{\sqrt{3}}A_8^\mu+
(1-T_W^2/3)^{1/2}B^\mu\right],\\ \nonumber
Z^\mu&=& C_W A_3^\mu - S_W\left[\frac{T_W}{\sqrt{3}}A_8^\mu+
(1-T_W^2/3)^{1/2}B^\mu\right],\\ \nonumber
Z'^\mu&=&-(1-T_W^2/3)^{1/2}A_8^\mu+\frac{T_W}{\sqrt{3}}B^\mu.
\end{eqnarray}
$S_W$ and $C_W$ are, respectively, the sine and cosine of the electroweak mixing angle  ($T_W=S_W/C_W$) defined by
$S_W=\sqrt{3}g'/\sqrt{3g^2+4g'^2}$. Also we can identify the $Y$
hypercharge associated with the SM gauge boson as:
\[Y^\mu=\left[\frac{T_W}{\sqrt{3}}A_8^\mu+ (1-T_W^2/3)^{1/2}B^\mu\right].\]
In the limit $\theta\longrightarrow 0$, $M_{Z}=M_{W^{\pm}}/C_W$, and
$Z_1^\mu=Z^\mu$ is the gauge boson of the SM. This limit
is obtained either by demanding $V\longrightarrow\infty$ or
$v'^2=v^2(C_W^2-S_W^2)$. In general $\theta$ may be
different from zero although it takes a very small value, determined from
phenomenology for each particular model.

\section{Conclusions}
In this paper we have studied the theory of
$SU(3)c\otimes SU(3)_L\otimes U(1)_X$ in detail. By restricting the
fermion field representations to particles without exotic electric charges
we end up with six different realistic models, two one family models and four models
for three families which are relatively new in the
literature, with two of them (models {\bf E} and {\bf F}) introduced
here for the first time, as far as we know.

If we allow for particles with exotic electric charges an infinite number
of models can be constructed, where the model in Ref. \cite{fp} is just
one of them.

The low energy predictions of the six models are not the
same. All of them have in common a new neutral current which mixes with
the SM neutral current which is also included as part of each
model (see Refs. \cite{long}-\cite{mps}).

The most remarkable result of our analysis is that, contrary to what is
stated in Ref. \cite{fp}, the 331 theory can be used to construct either
one family models or multi-family models, with the number of families
being a free number. Conspicuously enough are the existence of models
{\bf E} and {\bf F} for three families, where the three families are
treated different.

\section*{Acknowledgments}
This work was partially supported by BID and Colciencias in Colombia.

\end{document}